\documentclass[intlimits,twoside,a4paper]{article}

\usepackage{amsmath,amssymb}
\usepackage{graphicx}
\usepackage{wrapfig}

\usepackage[T2A]{fontenc}
\usepackage[cp1251]{inputenc}

\usepackage[eqsecnum]{cmpj2}

\issue{2013}{16}{4}{43602}
\doinumber{10.5488/CMP.16.43602}

\title[Lattice gas Lebwohl-Lasher model under confinement]%
{Phase behaviour of the confined lattice gas Lebwohl-Lasher model%
}
\author[N.G. Almarza, C. Martin, E. Lomba]{N.G. Almarza, C. Martin, E. Lomba}
\address{Instituto de Qu\'{\i}mica F\'{\i}sica Rocasolano, CSIC, Serrano 119,
  E-28006 Madrid, Spain}
\date{Received July 25, 2013}
\authorcopyright{N.G. Almarza, C. Martin, E. Lomba, 2013}

\begin{document}

\maketitle

\begin{abstract}
The phase behaviour of the Lebwohl-Lasher lattice gas model
(one of the simplest representations of a nematogenic fluid) confined in a
slab is investigated by means of extensive Monte Carlo
simulations. The model is known to yield a first order gas-liquid
transition in both the 2D and 3D limits, that is coupled with an
orientational order-disorder transition. This latter transition
happens to be first order in the 3D limit
and it shares some characteristic features with the continuous defect
mediated Berezinskii-Kosterlitz-Thouless transition in 2D.  In this
work we will analyze in detail the behaviour of this system taking
full advantage of the lattice nature of the model and the particular
symmetry of the interaction potential, which allows for the use of
efficient cluster algorithms.
\keywords Lebwohl-Lasher, orientational transitions, confinement, BKT transition
\pacs 64.70.-p, 64.60Cn, 61.20.Gy
\end{abstract}

\section{Introduction}
\looseness=-1The Lebwohl-Lasher (LL) model~\cite{PRA_1972_6_426} is known to be
one of the simplest systems that can reproduce the isotropic-nematic
transition, which is a key feature in the physics of liquid crystals,
ubiquitous materials in today's technology. It is  actually nothing but
a lattice version of the somewhat older continuum Maier-Saupe
model~\cite{ZNf_1959_14A_882}. Prof. Myroslav Holovko, whose 70th
birthday we are celebrating with this Festschrift, was one of the
first to study the solution of anisotropic integral equation
approaches to study the Maier-Saupe fluid and
its order-disorder transitions~\cite{JML_1999_82_161}, and he has recently published
a study in which the system is considered in terms of a field
theoretical approach~\cite{Kravtsiv2013}. In this work, as a tribute to
his numerous contributions to the field of statistical mechanics of
fluids and phase transitions, we will present a computer simulation
study of the phase behaviour of a Lebwohl-Lasher  lattice (LLL) gas under
confinement. This model is an extension of the previous work by
us~\cite{PRE_2010_82_011140} in which the nature of the orientational
transitions of the LL model was analyzed in depth, and
its close connection with the defect mediated
Berenzinskii-Kosterlitz-Thouless (BKT)
transition~\cite{JETP_1971_32_493,JPC_1972_5_L124} was investigated. Here, we add translational
degrees of freedom in terms of a lattice occupation variable. It is
worth to recall that the precise nature of the apparently ordered
phase in the two dimensional version of the LL model has been very
much discussed (see for instance reference~\cite{PRE_2008_78_051706}
and references therein).  In this paper we adopt the view of Reference
~\cite{PRE_2010_82_011140}, in the sense that these systems exhibit a transition
between an isotropic and a quasi-nematic phase endowed with
quasi-long range orientational order. We will
confirm the findings for the confined continuum Maier-Saupe fluid which
we investigated using computer simulations~\cite{PRE_2009_80_031501} and
extend the calculations to much larger system sizes that are
computationally unfeasible in continuum models. The coupling of the first order gas-liquid transition induced by
the net attraction between the spins and the BKT-like transition of
the confined fluid will be illustrated for various degrees of
confinement. This BKT-like orientational transition evolves gradually into  a weakly first order transition in
the 3D bulk limit~\cite{PRE_2001_64_051702}.

The rest of the paper is organized as follows. In the next section we
describe the model and the Monte Carlo methods used to study the
system. The last section is devoted to a presentation of the most
significant results  and conclusions.

\begin{table}[b]
\caption{Multicritical temperatures calculated using the GCWL
  methodology and equation~(\ref{g4c}).}
  \vspace{2ex}
\begin{center}
%
\begin{tabular}{|l|c|c|c|}
\hline
\multicolumn{1}{|c|}{$H$} & \multicolumn{1}{c}{$\rho_\textrm{tc}$} &
\multicolumn{1}{|c|}{$T_\textrm{tc}^*$} & \multicolumn{1}{|c|}{$\mu_\textrm{tc}/\epsilon$} \\
\hline\hline
   1   &  0.755 (11)   &  0.3723  &  --0.7643 (2)  \\
   2   &  0.741 (3)   &  0.5416  &  --0.6867 (2)  \\
   3   &  0.726 (4)   &  0.6139  &  --0.613  (2)  \\
   4   &  0.706 (2)   &  0.658   &  --0.549  (3)  \\
   8   &  0.717 (2)   &  0.767   &  --0.245  (7)  \\
\hline
\end{tabular}
\end{center}
\label{VCrit}
\end{table}

\section{Model and methods}
Our LL model is the simplest version of a nematogenic fluid, in which
every molecule interacts solely with molecules placed at neighbouring
sites of the lattice. Every lattice site can be either occupied or
empty, and this is controlled by an occupation number variable, $n_i$,
which can take the value 1 or 0 depending on whether there is a
molecule at the site $i$ or not. The total potential energy of the
system can thus be expressed as
\begin{equation}
U = -\varepsilon \sum_{\langle ij\rangle}n_in_j P_2({\bf s}_i{\bf s}_j),
\label{utot}
\end{equation}
where $\varepsilon$ is the coupling parameter that defines the energy
scale ($\varepsilon >0$)  and the reduced temperature $T^* =
k_\textrm{B}T/\varepsilon$ (where $k_\textrm{B}$ is Boltzmann's constant as usual), ${\bf s}_i$ and ${\bf s}_j$ are unit vectors
that describe the orientation of molecules $i$ and $j$, $P_2$ is the
second degree Legendre polynomial and finally $\langle ij \rangle$ indicates that
the summation is restricted to nearest neighbour (NN) pairs of
sites. Our model consists in a slab of cubic lattice, with periodic boundary
conditions in the $x$, $y$ directions and a width of $H$ sites along the $z$
direction as in the case of reference~\cite{PRE_2010_82_011140}, where
all sites were occupied. The model thus will have a total of $L\times
L\times H$ sites.
According to the  previous results~\cite{PRE_2005_71_046132},
 we expect to find a first-order (liquid-vapor-like) transition at low temperatures, and a continuous transition between an isotropic and a quasi-nematic phase at higher temperature. In the range of temperatures where the transition is continuous
 we have performed,
for each size $H$,
 simulations for a
series of $L$ values in order perform a finite size scaling analysis,
namely $L=10,20,30,40,60,80,90,100$.

\begin{figure}[h]
\centerline{
\includegraphics[width=0.45\textwidth,clip]{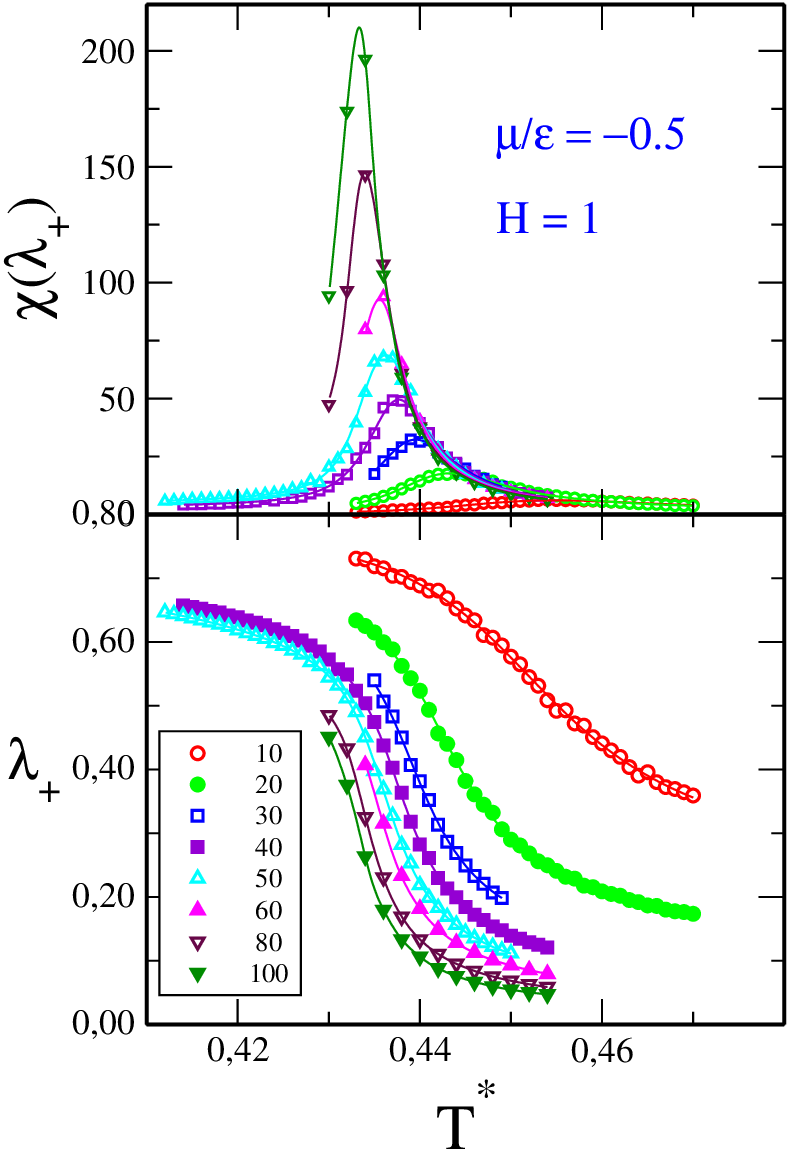}
\hspace{5mm}
\includegraphics[width=0.45\textwidth,clip]{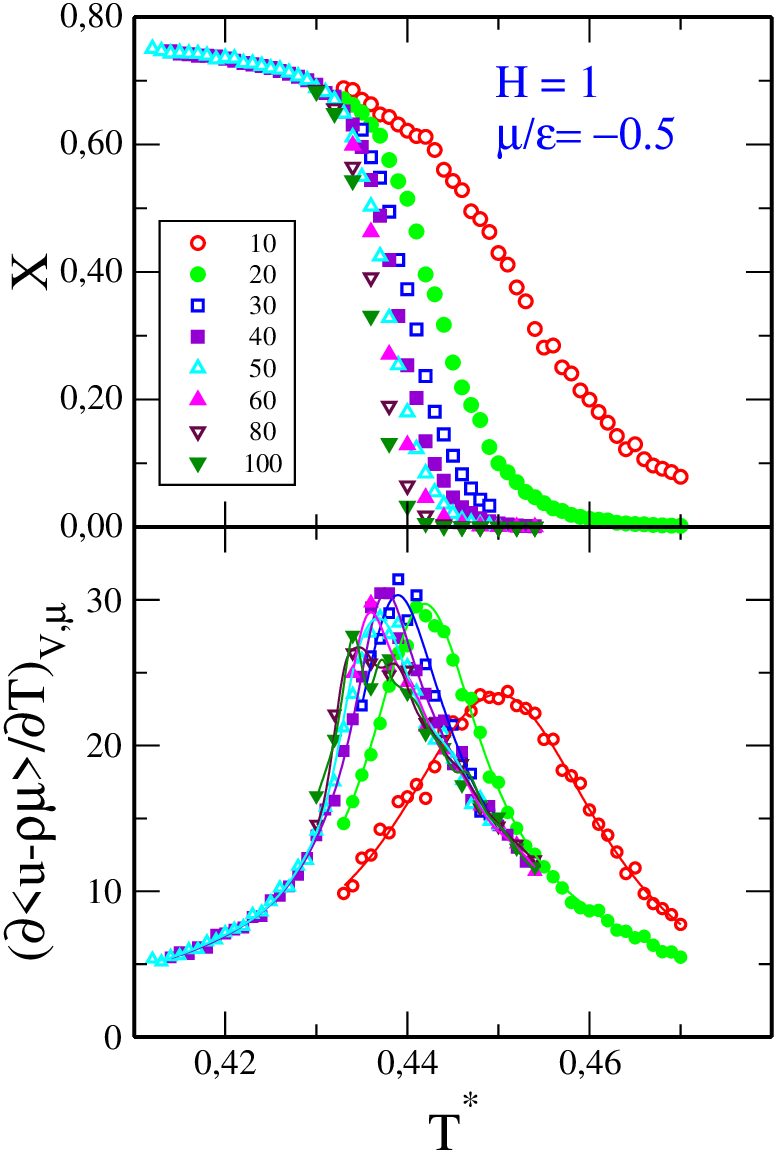}
}
\parbox[t]{0.5\textwidth}{
\caption{(Color online) Size dependence of the order parameter, $\lambda_+$, and
  susceptibility,  $\chi$ vs. temperature for a given chemical
  potential in the confined LL model.\label{VPS}}
}
\parbox[t]{0.5\textwidth}{
\caption{(Color online) Size dependence of the fraction of percolating clusters, $X$,
  and the grand canonical constant volume heat capacity vs. temperature
for the confined LL fluid.\label{PercMua}}
}
\end{figure}


As in the previous works,
for continuous transitions we have
performed Monte Carlo simulations using a combination of a local
update algorithm~\cite{PRE_2005_71_046132,PRL_1995_75_2887} and cluster
moves~\cite{PRE_2005_71_046132,PRL_1989_62_261,PRE_2011_83_041701} so as to minimize
critical slowdown effects when approaching the critical
temperature.
The temperature range in which  first-order transitions are expected
to occur was analysed by means of
simulations using
Grand-Canonical Wang-Landau (GCWL) methodology, following the prescription of
reference~\cite{PRE_2009_80_031501} for continuum models. We refer the
reader to our previous
works~\cite{PRE_2005_71_046132,PRE_2009_80_031501,PRE_2010_82_011140,JCP_2009_131_124506}
for technical details about the precise implementation of the methods. Once
more, the isotropic-quasi-nematic transition is monitored by means of
the largest eigenvalue, $\lambda_+$, of Saupe's tensor~\cite{ZNf_1964_19_161}
\begin{equation}
Q_{\alpha\beta} = \frac{1}{2N}\sum_{i=1}^N \left(3s_i^\alpha s_i^\beta
-\delta_{\alpha\beta}\right),
\label{saupe}
\end{equation}
where $N$ is the number of spins, and $\alpha$ and $\beta$ refer to
the $x$, $y$, $z$ components of the unit vector ${\bf s}_i$.
For a fixed value of the chemical potential, $\mu$,
we can define a size dependent pseudocritical temperature $T_\textrm{c}(\mu,L,H)$, in terms of
the fluctuation of the order parameter
\begin{equation}
\chi = N\left(\langle \lambda_+^2\rangle - \langle\lambda_+\rangle^2\right)/k_\textrm{B}T.
\label{chi}
\end{equation}
Using the maxima of the
$L$-dependent susceptibility as the estimate of $T_\textrm{c}(\mu,L)$,  as in previous works, we
have assumed a BKT-like finite size scaling of the
form~\cite{PRB_1992_46_662,PRB_2002_65_184405}
\begin{equation}
T_\textrm{c}(\mu,L) = T_\textrm{c}(\mu) + \frac{a_1}{(a_2+\mbox{ln} L)^2}\,.
\label{scal}
\end{equation}
In this way one can obtain the estimates of the BKT-transition
temperatures away from the first order transition.
Conversely, if the temperature is fixed,  we can use the same procedure to estimate
the pseudocritical chemical potential $\mu_\textrm{c}(T,L,H)$, and with an equation analogous to
equation (\ref{scal}) we compute $\mu_\textrm{c}(T)$.

\begin{figure}[h]
\centerline{
\includegraphics[width=0.435\textwidth,clip]{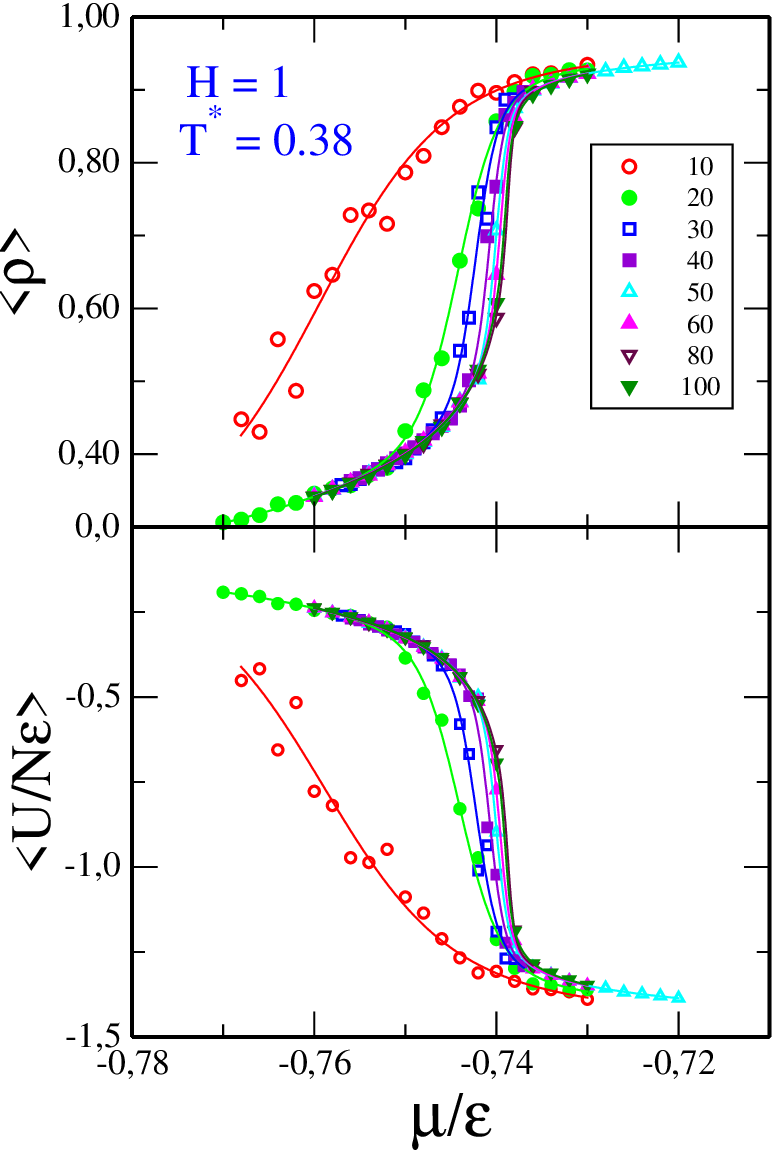}
\hspace{5mm}
\includegraphics[width=0.48\textwidth,clip]{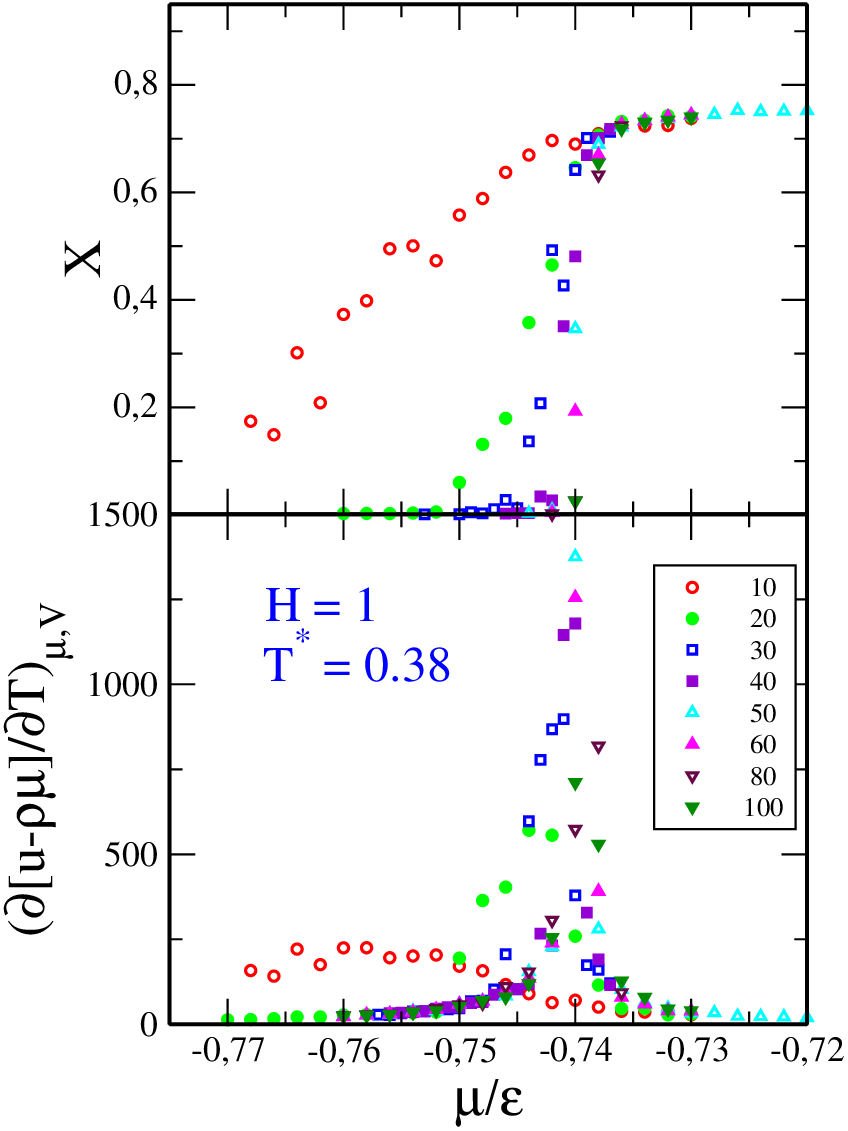}
}
\parbox[t]{0.5\textwidth}{
\caption{(Color online) Size dependence of density and potential energy vs. chemical
  potential for the confined LL fluid slightly above the multicritical temperature.\label{RhoU}}
}
\parbox[t]{0.5\textwidth}{
\caption{(Color online) Size dependence of the fraction of percolating clusters, $X$,
  and the grand canonical constant volume heat capacity vs. chemical
  potential
for the confined LL fluid slightly above the multicritical temperature.\label{PercMu}}
}
\end{figure}

In the case of the first order gas-liquid transition, we have carried
out computer simulations for different temperatures and systems sizes
using the GCWL method as follows:
\begin{itemize}
\item For a given $T$ and $L$, the Helmholtz energy is calculated for a
  series of densities in the fluid regime.
\item The possible phase transition is located by determining the value of
  the chemical potential $\mu_\textrm{e}(L,T)$ that maximizes the density
  fluctuations.
Paying attention to the grand canonical density distribution functions~\cite{PRE_2005_71_046132}
 $P(\rho;\mu_\textrm{e},T,L)$, and in particular to the dependence of their shape with $L$,
it is possible to probe the existence of the first order transition at the corresponding
temperature, and eventually to estimate the apparent coexistence densities of the
vapour and liquid phases, $\rho_\textrm{v}(L,T)$, and $\rho_\textrm{l}(L,T)$. These system-size
dependent densities can be used to extrapolate the results to the thermodynamic limit
($L\rightarrow \infty$).

\item The apparent multicritical temperature, $T_\textrm{tc}^{(L)} \equiv T_\textrm{tc}(L)$, is located
  with the aid of
  histogram reweighting techniques~\cite{PRL_1989_63_1195}. This
  multicritical temperature is defined as the temperature that
  fulfills~\cite{landau-binder_book_2005}
\begin{equation}
\frac{\langle\left[\delta\rho(T,L,\mu_\textrm{e})\right]^4\rangle}
{\langle\left[\delta\rho(T,L,\mu_\textrm{e})\right]^2\rangle^2} =
G_4^\textrm{c}\,,
\label{g4c}
\end{equation}
where $\delta\rho = \rho -\langle\rho\rangle$.  We have used, $G_4^\textrm{c}\approx
1.168$, i.e., the universal value of the amplitude ratio of the 2D
Ising model~\cite{JPhysA_1993_26_201}.
Then, we take the values $\mu_\textrm{tc}^{(L)}  \equiv \mu_\textrm{tc}(L) =
\mu_\textrm{e}(T_\textrm{tc}^{(L)},L)$, and
$\rho_\textrm{tc}^{(L)} \equiv \rho_\textrm{tc}(L) = \langle \rho(T_\textrm{tc}^{(L)}, L, \mu_\textrm{tc}^{(L)}) \rangle$ as estimates for the apparent (system-size dependent) multicritical quantities
$\mu_\textrm{tc}(L)$ and $\rho_\textrm{tc}(L)$.
\end{itemize}
Finally, the results for $\mu_\textrm{tc}(L)$,
$T_\textrm{tc}(L) $ and $\rho_\textrm{tc}(L) $ are fitted to second degree polynomials of $1/L$,
where the linear term is taken from the analysis of the related planar
Maier-Saupe model~\cite{PRE_2005_71_046132} and the quadratic term
accounts for the effects of confinement~\cite{PRE_2009_80_031501}. In
this way we can finally obtain the estimates for $L\rightarrow
\infty$.


\section{Results}
As mentioned before, calculations have been carried out for a series
of system sizes and values of $H=1, 2, 3, 4, 8$ and bulk 3D
system. The system can be thought of as either a slab in vacuum or a
fluid confined in an inert slit pore (wall-particle interactions
reduced to hard core exclusion).
In figure~\ref{VPS} we plot the order parameter and susceptibility results for the limiting case $H=1$ (a
2D LL fluid). As it is characteristic of BKT-like
transitions~\cite{PRE_2010_82_011140}, the order parameter, although
exhibits a clear jump across a transition temperature, tends to vanish
with an increasing sample size. On the other hand, the susceptibility
shows an apparent divergence, both at the transition temperature and
at temperatures below. The values of the fraction of percolating
clusters, $X$, shown in the upper graph of figure~\ref{PercMua}
exhibit a clear jump at the transition temperature. The
corresponding grand canonical constant volume heat capacity plotted in
the lower graph of the same figure (where $u=U/V$ is the potential
energy density), exhibits no size dependent
divergence, but most probably tends to  build up  a cusp singularity
characteristic of BKT-like transitions~\cite{PRB_1992_46_662}. Again,
similar results are obtained for temperatures/chemical potentials
above the transition temperature/chemical potential when analyzing the
results for other $H$ values, until in the limit of
$H\rightarrow\infty$ one recovers the bulk behaviour, i.e., a first
order phase transition as shown in reference~\cite{PRE_2010_82_011140}.

\begin{wrapfigure}{i}{0.5\textwidth}
\centerline{
\includegraphics[width=0.49\textwidth,clip]{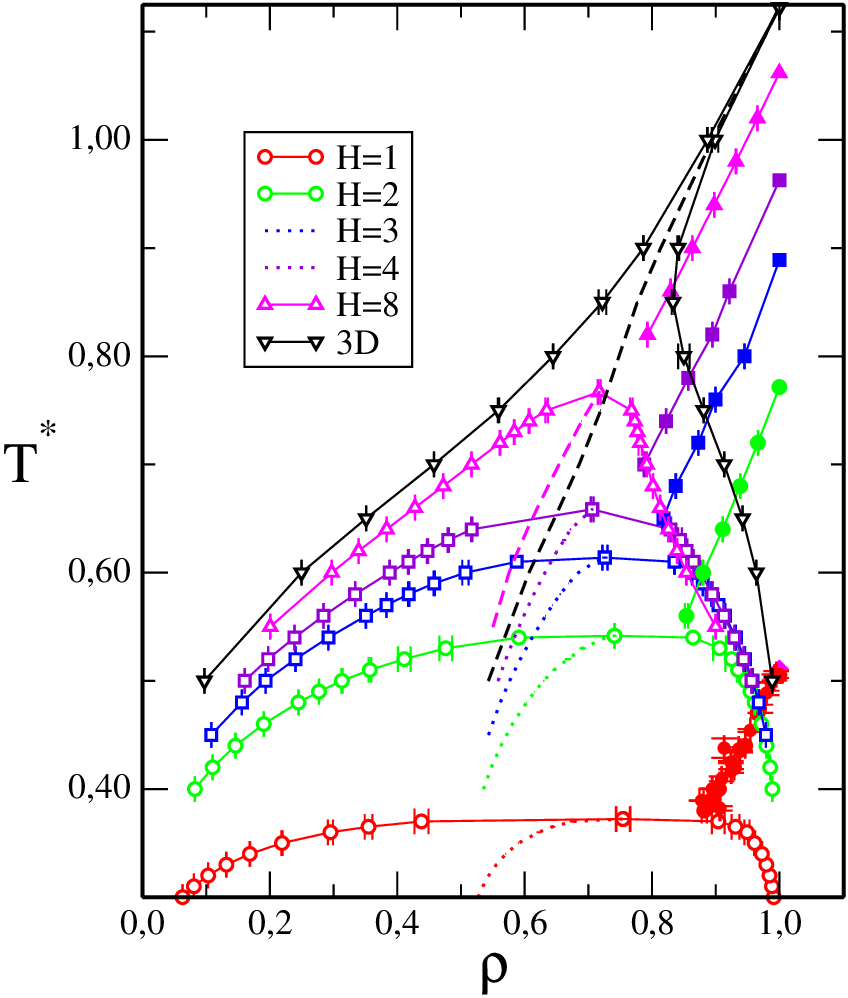}
}
\caption{(Color online) Phase diagram for the confined LLL model. Empty symbols
  correspond to the vapour-liquid equilibrium and solid ones to the
  continuous BKT-like transition. The rectilinear diameter lines are
  denoted by discontinuous curves. \protect\label{DBMx}
}
\end{wrapfigure}

For finite pore widths below a certain multicritical temperature
$T_\textrm{tc}$, one encounters first order transitions.
This is illustrated in
figure~\ref{RhoU} and in figure~\ref{PercMu}, where one can appreciate
large variations of the density and the energy with the chemical potential
 correlated with the BKT-like transition for a temperature slightly above $T_\textrm{tc}$.
 These multicritical temperatures (the highest $T$ at which
the isotropic-quasinematic transition is coupled with a first order
gas-liquid transition) are estimated using the approach indicated in
the previous section, leading to the results of table~\ref{VCrit}.

The $T-\rho$ phase diagram for the LLL model under varying degrees of
confinement is presented in figure~\ref{DBMx}. We see how the
relatively flat (Ising 2D) behaviour of the vapour-liquid curve for
$H=1$ gradually
 evolves into the shape of the bulk 3D LL
isotropic-nematic transition in which the orientational transition is
fully coupled to the vapour-liquid transition for all temperatures. If
one tries to correlate the value of the multicritical temperatures
with the pore, we again find that the behaviour deviates somewhat
from the Kelvin-like scaling $T_\textrm{c}(\text{bulk}) - T_\textrm{c}(H) \propto 1/H$ as it was
the case in the confined Maier-Saupe continuum case.
Due to the presence of the order-disorder transition, the multicritical
points cannot be directly mapped into vapour-liquid critical points,
in particular, as the width of the pore increases. As a matter of fact,
the $H=1$ phase diagram might also appear to be compatible with the
presence of a critical endpoint, instead of a tricritical point, but
on the basis of the behaviour for $H > 1$ one can think that most
likely the BKT line for $H=1$ (solid red circles) bends backwards as
$T$ is lowered to
hit the vapour-liquid equilibrium curve right at the top (the VL
critical point). As $H$ is increased this behaviour is more apparent.

In summary, we have presented a detailed analysis of the phase
behaviour of the confined Lebwohl-Lasher lattice model. We have shown
how the confinement transforms the first order isotropic-nematic transition
of the bulk 3D LL system into a continuum BKT transition between an
isotropic phase and a phase with local orientational order
(quasinematic), which develops into a first order vapour-liquid
transition below a given multicritical temperature. These findings
confirm and extend those for the continuum Maier-Saupe model with a
more reliable finite size scaling analysis.

\section*{Acknowledgements}
The authors gratefully acknowledge the support from the Direcci\'on
General de Investigaci\'on Cient\'{\i}fica  y T\'ecnica under Grant
No. FIS2010-15502 and from the Direcci\'on General de
Universidades e Investigaci\'on de la Comunidad de Madrid under
Grant No. S2009/ESP/1691 and Program MODELICO-CM.



\begin{thebibliography}{99}


\bibitem{PRA_1972_6_426}
Lebwohl P.A., Lasher G., Phys. Rev. A, 1972, \textbf{6}, 426; \doi{10.1103/PhysRevA.6.426}.

\bibitem{ZNf_1959_14A_882}
Maier W., Saupe A., Z. Naturforsch., 1959, \textbf{14A}, 882.

\bibitem{JML_1999_82_161}
Holovko M., Sokolovska T., J. Mol. Liq., 1999, \textbf{82}, 161; \doi{10.1016/S0167-7322(99)00098-7}.

\bibitem{Kravtsiv2013}
Kravtsiv I., Holovko M., di~Caprio D., Mol. Phys., 2013, \textbf{111}, 1023; \doi{10.1080/00268976.2012.762615}.

\bibitem{PRE_2010_82_011140}
Almarza N.G., Martin C., Lomba E., Phys. Rev. E, 2010, \textbf{82}, 011140; \doi{10.1103/PhysRevE.82.011140}.

\bibitem{JETP_1971_32_493}
Berezinskii V., Sov. Phys.-JETP, 1971, \textbf{32}, 493.

\bibitem{JPC_1972_5_L124}
Kosterlitz J.M., Thouless D.J., J. Phys. C: Solid State Phys., 1972,
  \textbf{5}, L124; \doi{10.1088/0022-3719/5/11/002}.

\bibitem{PRE_2008_78_051706}
Paredes R., Farinas-S\'{a}nchez A.I., Botet R., Phys. Rev. E, 2008,
  \textbf{78}, 051706; \doi{10.1103/PhysRevE.78.051706}.

\bibitem{PRE_2009_80_031501}
Almarza N.G., Martin C., Lomba E., Phys. Rev. E, 2009, \textbf{80}, 031501; \doi{10.1103/PhysRevE.80.031501}.

\bibitem{PRE_2001_64_051702}
Bates M.A., Phys. Rev. E, 2001, \textbf{64}, 051702; \doi{10.1103/PhysRevE.64.051702}.

\bibitem{PRE_2005_71_046132}
Lomba E., Mart\'{\i}n C., Almarza N.G., Lado F., Phys. Rev. E, 2005,
  \textbf{71}, 046132; \doi{10.1103/PhysRevE.71.046132}.

\bibitem{PRL_1995_75_2887}
Nijmeijer M.J.P., Weis J.J., Phys. Rev. Lett., 1995, \textbf{75}, 2887; \doi{10.1103/PhysRevLett.75.2887}.

\bibitem{PRL_1989_62_261}
Wolff U., Phys. Rev. Lett., 1989, \textbf{62}, 361; \doi{10.1103/PhysRevLett.62.361}.

\bibitem{PRE_2011_83_041701}
Marguta R.G, Mart\'{\i}nez-Rat\'on Y., Almarza N.G., Velasco E., Phys. Rev. E, 2011,
\textbf{83}, 041701; \\ \doi{10.1103/PhysRevE.83.041701}.

\bibitem{JCP_2009_131_124506}
Almarza N.G., Capit\'an J.A., Cuesta J.A., Lomba E., J. Chem. Phys., 2009, \textbf{131}, 124506; \doi{10.1063/1.3223999}.


\bibitem{ZNf_1964_19_161}
Saupe A., Z. Naturforsch., 1964, \textbf{19}, 161.

\bibitem{PRB_1992_46_662}
Kunz H., Zumbach G., Phys. Rev. B, 1992, \textbf{46}, 662; \doi{10.1103/PhysRevB.46.662}.

\bibitem{PRB_2002_65_184405}
Tomita Y., Okabe Y., Phys. Rev. B, 2002, \textbf{65}, 184405; \doi{10.1103/PhysRevB.65.184405}.

\bibitem{PRE_1999_59_2168}
Weber H., Paul W., Binder K., Phys. Rev., 1999, \textbf{59}, 2168; \doi{	 10.1103/PhysRevE.59.2168}.

\bibitem{PRL_1989_63_1195}
Ferrenberg A.M., Swendsen R.H., Phys. Rev. Lett., 1989, \textbf{63}, 1195; \doi{10.1103/PhysRevLett.63.1195}.

\bibitem{landau-binder_book_2005}
Landau D.P., Binder K., A Guide to Monte Carlo Simulations in Statistical
  Physics, Cambridge University, Cambridge, 2005.

\bibitem{JPhysA_1993_26_201}
Kamieniarz G., Blote H., J. Phys. A: Math. Gen., 1993, \textbf{26}, 201; \doi{10.1088/0305-4470/26/2/009}.

\end{thebibliography}

\ukrainianpart
\title{Фазова поведінка просторово обмеженої граткової моделі Лебволя-Лашера%
}
\author{Н. Альмарца, К. Мартін, Е. Ломба}
\address{Інститут фізичної хімії, Серрано, E--28006 Мадрид, Іспанія}
\makeukrtitle
\begin{abstract}
\tolerance=3000%
Фазова поведінка  граткової моделі Лебволя-Лашера   (одне з
найпростіших представлень нематогенного плину), що знаходиться між
двома площинами, досліджується за допомогою екстенсивних симуляцій
Монте Карло. Відомо, що модель дає фазовий перехід газ-рідина
першого роду в двовимірній і тривимірній границях, який
пов'язаний з орієнтаційним переходом лад-безлад. Останній є переходом першого роду в тривимірній границі і має деякі спільні
характеристичні риси з неперервним  дефектно опосередкованим
переходом Березінського-Костерліца-Таулесса у двовимірному випадку.
В цій роботі ми детально аналізуємо поведінку цієї системи,
використовуючи всі переваги ґраткової природи моделі і  симетрії
потенцалу взаємодії, які дають змогу використати ефективні
кластерні алгоритми.
\keywords Лебволь-Лашер, орієнтаційні переходи, конфайнмент, перехід
BKT
\end{abstract}

\end{document}